\documentclass[twocolumn,showpacs,preprintnumbers,amsmath,amssymb]{revtex4}

\usepackage{graphicx}
\usepackage{dcolumn}
\usepackage{bm}
\usepackage{multirow}
\def\id{{\mathchoice {\rm 1\mskip-4mu l} {\rm 1\mskip-4mu l} {\rm 1\mskip-4.5mu l} {\rm 1\mskip-5mu l}}}

\begin{document}

\preprint{APS/123-QED}

\title{Solid-state NMR three-qubit homonuclear system for quantum information processing: control and characterization}

\author{Jonathan Baugh}
\email{baugh@iqc.ca}
\author{Osama Moussa}
\author{Colm A. Ryan}
\author{Raymond Laflamme}
\email{rlaflamme@iqc.ca}
\homepage{http://www.iqc.ca}
\altaffiliation[Also at ]{Perimeter Institute for Theoretical Physics, Waterloo, ON}
\affiliation{%
Institute for Quantum Computing, University of Waterloo, Waterloo, Ontario N2L 3G1\\
}%
\author{Chandrasekhar Ramanathan}
\author{Timothy F. Havel}
\author{David G. Cory}
\affiliation{
Department of Nuclear Science and Engineering, Massachusetts Institute of Technology, Cambridge, MA 02139\\
}%

\date{\today}

\begin{abstract}
A three-qubit $^{13}$C solid-state nuclear magnetic resonance
(NMR) system for quantum information processing, based on the
malonic acid molecule, is used to demonstrate high-fidelity
universal quantum control via strongly modulating radio-frequency
pulses. This control is achieved in the strong-coupling regime, in
which the timescales of selective qubit addressing and of
two-qubit interactions are comparable.  State evolutions under the
internal Hamiltonian in this regime are significantly more
complex, in general, than those of typical liquid-state NMR
systems. Moreover, the transformations generated by the strongly
modulating pulses are shown to be robust against the types of
ensemble inhomogeneity that dominate in the employed molecular
crystal system.  The secondary focus of the paper is upon detailed
characterization of the malonic acid system. The internal
Hamiltonian of the qubits is determined through spectral
simulation. A pseudopure state preparation protocol is extended to
make a precise measurement of the dephasing rate of a
three-quantum coherence state under residual dipolar interactions.
The spectrum of intermolecular $^{13}$C-$^{13}$C dipolar fields in
the crystal is simulated, and the results compared with
single-quantum dephasing data obtained using appropriate
refocusing sequences. We conclude that solid-state NMR systems
tailored for quantum information processing have excellent
potential for extending the investigations begun in liquid-state
systems to greater numbers of qubits.
\end{abstract}

\maketitle

\section{Introduction}

\indent Quantum information processing (QIP) aims to achieve the
ultimate in computational power from physical systems by
exploiting their quantum nature \cite{NC00a}.  Nuclear magnetic
resonance (NMR)-based QIP has been successfully implemented in
liquid-state ensemble systems of up to $7$ qubits
\cite{CPH98a,VSBY01a,YSV+99a,CLK+00a,KLMT00a,DBK04a}. Universal
quantum control is achieved through application of external
radio-frequency (RF) fields on or near resonance with spin
transitions of a set of separately addressable, coupled spins.
State initialization (to a fiducial state such as $|00..0\rangle$)
is effectively achieved in these systems by the preparation of
pseudopure states \cite{CPH98a,KCL98a}.  Pseudopure states have
recently been demonstrated in a 12-qubit liquid system
\cite{NMR+05a} and in a 12-spin liquid-crystal system
\cite{LK05a}. A hallmark of control in liquid-state systems is a
separation of timescales between (faster) qubit addressability and
(slower) two-qubit coupling gates, typically by an order of
magnitude or more \cite{CLK+00a}. For the homonuclear subsystems,
this corresponds to the weak-coupling regime, in which the
relative Zeeman shifts in the qubit Hamiltonian are significantly
larger than the \textit{J}-couplings \cite{Abr61a,EB87a}.  In this
regime, the evolution due to spin interactions is predominantly of
the controlled-phase form. In this work, we examine NMR-based QIP
implemented in a \emph{solid-state} homonuclear system, in which
the qubit Hamiltonian is no longer in the weak-coupling regime.  A
solid-state NMR architecture is attractive due to several key
properties \cite{CLK+00a}: (1) nuclear spin states have been
purified to near-unity polarizations in solids \cite{AG82a}; (2)
intrinsic decoherence times can be much longer, and two-qubit gate
times much shorter, than those in the liquid state; (3) the qubit
spins can be brought into well-controlled contact with a
thermal-bath of bulk spins, enabling entropy-reducing operations
such as algorithmic cooling \cite{SV90a,BMR+05a} and quantum error
correction \cite{KLZ98a,LMPZ96a} to be carried out.  The system we
will describe is specially tailored so that the ensemble
description of the system is--to a good approximation--analogous
to that of liquid-state NMR-QIP, and therefore the general aspects
of control and measurement are the same \cite{CLK+00a}.  A similar
three qubit system using single-crystal glycine was first explored
by Leskowitz et. al. \cite{LOGM03a}, in which the homonuclear
two-carbon system was approximately weakly-coupled. However, we
will demonstrate that universal, coherent control can be
implemented in the strong-coupling regime: the regime in which the
timescales of qubit addressing and qubit coupling are comparable.
In this regime, the transverse spin operator terms from
qubit-qubit dipolar couplings are less suppressed by the relative
Zeeman shifts.  These residual 'flip-flop' terms
$\sigma^{j}_{+}\sigma^{k}_{-}+\sigma^{j}_{-}\sigma^{k}_{+}$ (where
$\sigma_{\pm}=\sigma_{x}\pm i\sigma_{y}$ and
$\{\sigma_{x},\sigma_{y}\}$ are Pauli matrices) render the state
evolutions more complex, in general, than those of weakly-coupled
spin systems. We show that strongly modulating pulses
\cite{FPB+02a} succeed in controlling the solid-state QIP system,
with generality and with high-fidelity. Moreover, the pulses
provide significant robustness for the desired transformations
against the ensemble inhomogeneities that are typical of
solid-state NMR systems.  This latter property is of great
importance to the practical application of quantum algorithms in
such systems, and we present here a first step in its study. This
control methodology is a key ingredient in the realization of
solid-state NMR-QIP testbed devices, but could also extend to many
other potential systems for quantum information processing.  We
remark that liquid-crystalline NMR-QIP implementations
\cite{DBK04a,LK04a} represent a coupling regime that is typically
intermediate between the strong- and weak-coupling cases. Strongly
modulating pulses would therefore also be an appropriate
means for controlling such systems universally. \\
\indent The secondary focus of this paper is to characterize, in
detail, the employed three-qubit system based on the malonic acid
molecule.  This includes characterization of the dominant ensemble
inhomogeneities arising both through linear and bi-linear terms in
the Hamiltonian. In tailoring the present system, we have used
dilution of the qubit molecules as a means for reaching an
approximate ensemble description in which processor molecules are,
ideally, non-interacting and reside in identical environments.
However, the need for a macroscopic number of spins to generate
observable NMR signals through the usual inductive detection both
limits the degree of dilution and requires a macroscopic sample.
The former results in perturbing intermolecular dipolar fields,
and the latter typically yields significant distortions of the
applied magnetic fields over the sample volume, namely, of the
static magnetic field (due to susceptibility/shape effects) and of
the RF amplitude (due to the RF coil geometry). In this paper, we
address the robustness of strongly modulating pulse control to
dispersion of Zeeman shifts and of RF amplitudes. We also
characterize in detail the intermolecular dipolar environment in
the present
system, both theoretically and experimentally.\\
\indent The paper is organized as follows: section~\ref{sec:sec3}
reviews strongly modulating RF pulses as a means of achieving
universal quantum control, and numerical results for an example
pulse are discussed; in section~\ref{sec:sec2}, the dilute
$^{13}$C malonic acid system is first characterized;
section~\ref{sec:sec4} demonstrates the preparation of a
pseudopure state (as a benchmark for control) and analyzes the
results; section~\ref{subsec:sec4c} treats an application of the
pseudopure state protocol: precise measurement of the
triple-quantum dipolar dephasing rate, and comparison to
single-quantum dephasing rates; in section~\ref{sec:sec5},
simulated and experimental data are presented that explore the
effect of intermolecular dipolar couplings on qubit coherence
times; in section~\ref{subsec:sec5b}, a multiple-pulse refocusing
sequence is used in order to compare appropriate experimental
quantities with the intermolecular dipolar simulations, and as a
first step in testing the attainability of long ensemble coherence
times in this system; the overall results are discussed in
section~\ref{sec:sec6}, in the context of assessing the future
goals and potential of solid-state NMR-QIP.

\section{Universal Control: Strongly modulating pulses}\label{sec:sec3}
\begin{figure}
\includegraphics[height=12cm]{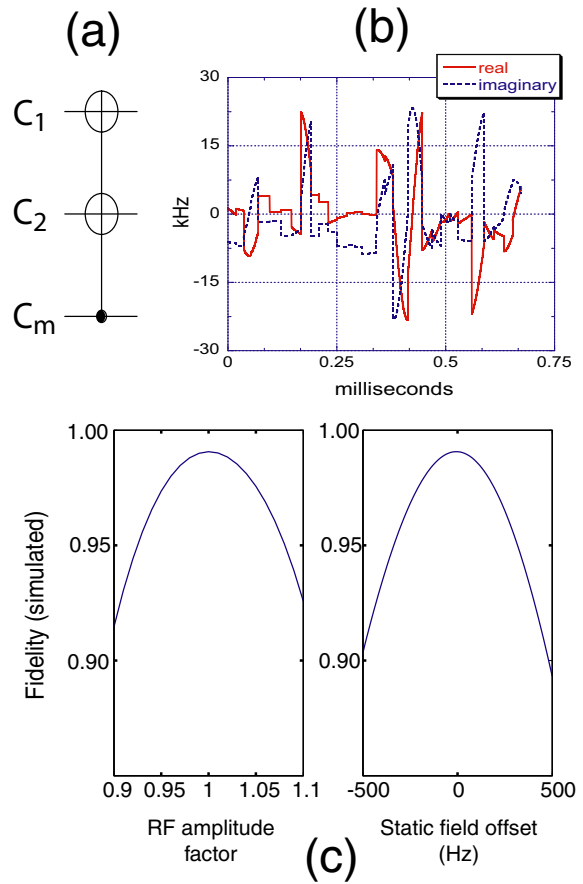}
\caption{Example of a modulated RF pulse solution for generating a
three-qubit gate in the malonic acid system. (a) quantum network
representation of the gate, which is a
controlled-(not$\otimes$not), with $C_m$ as the control bit and
$C_1$ and $C_2$ as target bits. (b) the quadrature components of
the RF modulation waveform, labelled 'real' and 'imaginary'. (c)
Simulated fidelity of the transformation as a function of (left)
scaling of the RF amplitude and (right) offset of the static
magnetic field. \label{fig:fig1}}
\end{figure}

\indent Numerically optimized 'strongly modulating' pulses were
previously introduced as a means of implementing fast,
high-fidelity unitary gate operations in the context of
liquid-state NMR-QIP \cite{FPB+02a}. The aim is to construct an
arbitrary modulated RF waveform that, when applied to the system,
generates a desired effective Hamiltonian corresponding to a
particular quantum gate. This is accomplished numerically by
minimizing the distance between the actual and the desired unitary
transformations using a simplex search algorithm. Gate fidelities
are defined by the expression:
\begin{equation}
F=\sum_{\mu}p_{\mu}|Tr(U^{\dag}_{des}U^{\mu}_{calc})/N|^{2}
\end{equation}
where $N$ is the dimension of the Hilbert space, $U_{des}$ is the
desired unitary and $U_{calc}$ is the unitary calculated for the
evolution of the system under the modulated RF pulse. Here,
$p_{\mu}$ is a normalized empirical distribution over an
inhomogeneous parameter (or parameters) of the ensemble, typically
the RF amplitude. The expression for $F$ corresponds to an average
fidelity over all possible input states \cite{FPB+02a}.  The
number of parameters the algorithm must search over is made
minimal by requiring the modulating waveform to consist of a
series of constant amplitude and frequency periods, so that each
period has a time-independent Hamiltonian in a particular rotating
reference frame \cite{FPB+02a}. Modulation pulses with ideal
(simulated) fidelities of order $F\sim 99\%$ are readily found for
liquid-state NMR-QIP systems with up to 6 qubits \cite{NMR+05a,FPB+02a}.\\
\indent It is observed empirically that good modulating pulse
solutions tend to have the average RF amplitude $\nu_{RF}\sim
|\mathcal{H}|$, where $|\mathcal{H}|$ is the magnitude of the
internal qubit Hamiltonian. It is precisely this strong driving
regime in which analytical techniques (e.g. perturbation theory)
for calculating dynamics break down, yet the system generally has
the broadest (and most rapid) access to the manifold of allowed
states. It is also the regime in which all accessible spin
transitions are excited, so that refocusing of unwanted
interactions becomes possible, even to the extent that the
dephasing effects of ensemble inhomogeneities may be partially
suppressed \cite{BEH+04a,R02a}. \\
\indent An example of a strongly modulating RF pulse is shown in
Fig.~\ref{fig:fig1}.  The pulse is tailored to generate a
three-qubit controlled-(not$\otimes$not) gate in the
strongly-coupled malonic acid system to be described in the next
section. The average RF amplitude is $9.4$ kHz, whereas the
magnitude of the internal $^{13}$C Hamiltonian (parameters listed
in Table~\ref{tab:table1}, next section) is $|\mathcal{H}|\simeq
7.3$ kHz. Part (c) of the figure demonstrates the robustness of
the calculated unitary over the dominant inhomogeneous Hamiltonian
parameters of the ensemble, namely, scaling of the RF amplitude
and offset of the static field. This pulse was optimized over a
5-point probability distribution $p_{\mu}$ of RF amplitude scaling
(corresponding to that measured in our RF coil), centered on
unity, with standard deviation $\sigma(p_{\mu})\simeq 6\%$. The
fact that the unitary fidelity is $>90\%$ over a $1$ kHz range of
static field offset demonstrates the ability of the modulated
pulse to effectively refocus evolution under Zeeman shift
dispersion (no Zeeman distribution was used in the optimization).
An ideal fidelity $\sim\!98\%$ was obtained here, however, even
greater precision is likely required for successful general
implementation of quantum
algorithms. We consider this a first step that can be significantly improved upon in future work. \\
\indent The strongly modulating pulse methodology generates fast
pulses, relative to traditional selective pulse methods, to
implement unitary gates. Although the example pulse of
Fig.~\ref{fig:fig1} has a duration $\simeq 700 \mu$s, we have
found pulse solutions for the same gate (fidelities $>90\%$) as
short as $\simeq 450\mu$s. This compares well with the same gate
carried out as two separate controlled-not gates implemented using
standard low-amplitude selective $\pi/2$ pulses and $\pi/2$
controlled-phase evolutions; we estimate such a gate would require
at least $\sim 1.25$ ms to implement.  Optimal control theory has
been used by other workers to design numerical procedures for
constructing near time optimal state transformation pulses
\cite{KBG01a}.

\section{Characterization I: Ensemble Hamiltonian}\label{sec:sec2}
\subsection{Dilute 13C-labeled single-crystal malonic acid}
\begin{figure}
\includegraphics[height=4.5cm]{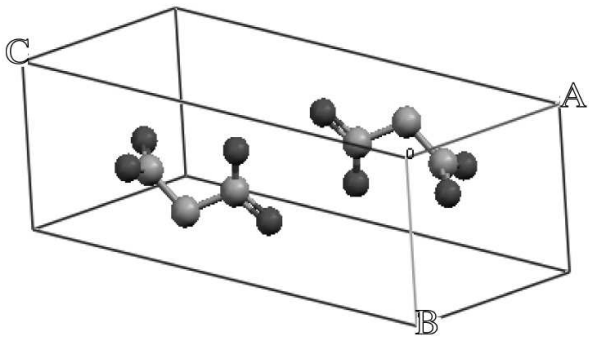}
\caption{Unit cell of malonic acid, with crystal axes indicated.
The two molecules are related by inversion symmetry, and are
therefore magnetically equivalent (P-1 space group). The light
(dark) atoms are carbon (oxygen), and the hydrogen atoms are not
shown. $\vec{A}=5.156$\AA, $\vec{B}=5.341$\AA\, and
$\vec{C}=8.407$\AA \cite{JRS94a}.\label{fig:fig2}}
\end{figure}

\begin{figure}
\includegraphics[height=7.5cm]{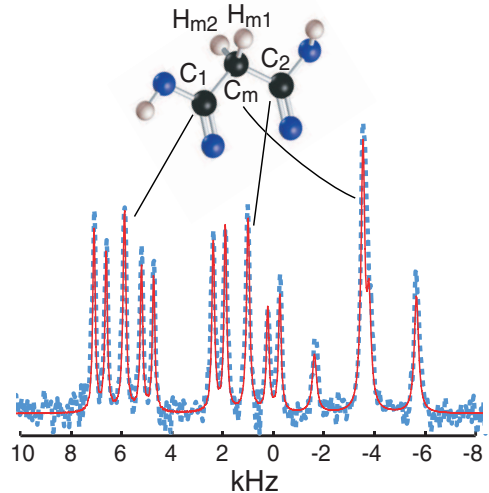}
\caption{$^{13}$C-NMR spectrum (dotted line) and spectral fitting
results (solid line) for $3.2\%$ $^{13}$C-labelled malonic acid in
the crystal orientation used throughout.  The fitting procedure
adjusts the simulated Hamiltonian to match the spectrum, and uses
Lorentzian line broadening to fit the $(T_2^{*})^{-1}$ depashing
rates. Proton decoupling was applied using a TPPM sequence
\cite{BRA+95a} at an RF power of $250$ kHz. Note the most intense
peaks, in the center of each multiplet, correspond to the natural
abundance ($1.1\%$) $^{13}$C. The uneven heights of the
labelled-molecule peaks indicate that the relative Zeeman shifts
are insufficient to fully truncate the strong form of the
intramolecular dipolar couplings; hence, the eigenstates of the
system are near, but not equal to, the usual computational basis
states. \label{fig:fig3}}
\end{figure}

\indent The system under study is a single crystal of malonic acid
grown from aqueous solution with a dilute fraction of fully
$^{13}$C-labelled molecules.  The main crystal used herein has
dimensions of $4\!\times\!1.5\!\times\!1.5$ mm$^{3}$ and a
labelled molecule fraction of $3.2\%$ (a similar crystal with
$1.6\%$ dilution was used in the experiments of
section~\ref{subsec:sec5b}). For the malonic acid molecule in the
solid, the three carbon nuclei have distinct anisotropic chemical
shielding tensors so that when placed in a large static magnetic
field, crystal orientations can be found for which each carbon is
separately addressable in frequency. Protons, of which there are
four per molecule (and are $100\%$ abundant in the crystal), can
be used to cross-polarize the carbon spins, and are otherwise
decoupled from the $^{13}$C system. Figure ~\ref{fig:fig2} shows
the malonic acid unit cell as determined by x-ray diffraction
\cite{JRS94a}. The space group is P-1 so that the two molecules in
the unit cell are related by inversion symmetry, and are therefore
magnetically equivalent. The crystal orientation is chosen to
maximize the intramolecular $^{13}$C-$^{13}$C dipolar couplings
and the relative $^{13}$C Zeeman shifts.  The full spin
Hamiltonian of the system is
\begin{equation}
\mathcal{H}(t)=\mathcal{H}_{C}+\mathcal{H}_{H}+\mathcal{H}_{CH}+\mathcal{H}_{RF}(t),
\end{equation}
where $\mathcal{H}_{C}$ and $\mathcal{H}_{H}$ are the $^{13}$C and
$^{1}$H Hamiltonians, $\mathcal{H}_{CH}$ is the interspecies
coupling Hamiltonian, and $\mathcal{H}_{RF}(t)$ is the
time-dependent Hamiltonian of the external radio-frequency field.
In many experiments, $\mathcal{H}_{RF}(t)$ includes a strong field
resonant with the $^{1}$H spins so that $\mathcal{H}_{CH}$ is
effectively removed from the Hamiltonian. The $^{13}$C Hamiltonian
can be decomposed as
\begin{equation}
\mathcal{H}_{C}=\mathcal{H}_{CZ}+\mathcal{H}^{intra}_{CD}+\mathcal{H}^{inter}_{CD},
\end{equation}
where the terms on the right side, from left to right, are the
Zeeman, the intramolecular dipolar, and the intermolecular dipolar
terms.  The Zeeman and intramolecular dipolar terms dominate the
natural $^{13}$C Hamiltonian in the dilute $^{13}$C crystal, and
will be used (along with the RF) in the construction of quantum
gates. The intermolecular couplings act as perturbations on these
terms, and their effects will be examined in section V. Denoting
single-spin Pauli matrices as
$X=\sigma_{x}\quad,Y=\sigma_{y}\quad,Z=\sigma_{z}$, the Zeeman and
intramolecular dipolar terms may be expressed as
\begin{align}
&\mathcal{H}_{CZ}=\sum^{3}_{j=1}\frac{\nu_{j}}{2}Z^{j}\\
&\mathcal{H}^{intra}_{CD}=\sum_{m<n\leq
3}\frac{d_{mn}}{4}(2Z^{m}Z^{n}-Y^{m}Y^{n}-X^{m}X^{n}),\label{eq:dip}
\end{align}
where $\nu_{j}$ are the rotating-frame Zeeman frequencies and
$d_{mn}$ are the dipolar coupling strengths. Table I lists these
parameters, as obtained from fitting the $^{13}$C spectrum of
Figure~\ref{fig:fig3}, for the crystal orientation used
throughout. It also lists the free-induction dephasing times,
$T_2^{*}$; the corresponding rates $(T_2^{*})^{-1}$ provide a
measure of the degree of ensemble inhomogeneity. It will be seen
in section V. that the dominant contribution to these rates is
Zeeman shift dispersion.

\begin{table}
\caption{\label{tab:table1} Zeeman shifts and intramolecular
dipolar couplings of the $^{13}$C Hamiltonian, columns 1-3 from
left; dipolar couplings involving the methylene protons, columns
4-5; $^{13}$C free-induction dephasing times ($T_2^{*}$) and
spin-lattice relaxation times ($T_1$). Values in columns 1-5 are
in kHz.}
\begin{ruledtabular}
\begin{tabular}{cccccccc}
 &$C_1$&$C_2$&$C_m$&$H_{m1}$&$H_{m2}$&$T_2^{*}$(ms)&$T_1$(s)\\
\hline
$C_1$&$5.893$&$0.227$&$0.935$&$-1.5$&$2.0$&$2.4$& $160$\\
$C_2$&  &$1.057$&$1.070$&$1.4$&$1.0$&$2.0$& $325$\\
$C_m$&  & &$-3.445$&$-18.7$&$-0.9$&$1.5$& $315$\\
\end{tabular}
\end{ruledtabular}
\end{table}

\subsection{Experimental setup} \indent All experiments were carried out at room temperature on a Bruker Avance NMR
spectrometer operating at a field of $7.1$ T, and home-built
dual-channel RF probehead. The sample coil had an inner diameter
of $3$ mm, and the typical $\pi/2$ 'hard' pulse lengths were $1.25
\mu$s and $0.75 \mu$s for carbon and hydrogen, respectively.

\begin{figure}
\includegraphics[height=4.25cm]{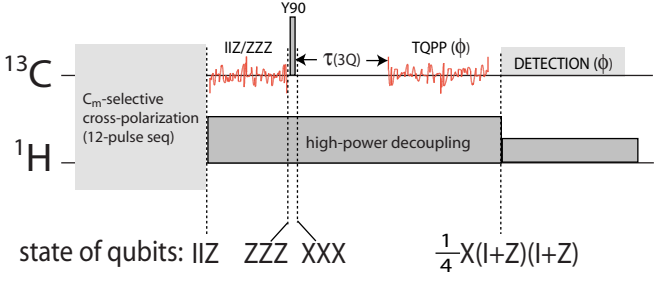}
\caption{Schematic pulse sequence for pseudopure state preparation
using strongly modulating pulses.  The states of the qubits are
indicated in product operator format after each transformation.
The state $\id^{1}\id^{2}Z^{m}$ is denoted as 'IIZ', for example.
Strongly modulating pulses are represented by waveforms labelled
'IIZ/ZZZ' and 'TQPP', whose durations are $0.54$ ms and $1.10$ ms,
respectively. A collective $\pi/2$ rotation about the
rotating-frame $\hat{y}$-axis is labelled 'Y90'. The angle $\phi$
represents a six-fold phase-shifting of the RF to filter out all
coherences except the 3Q coherence. $\tau(3Q)$ indicates a
variable delay subsequent to the creation of the 3Q state that is
used to measure that state's decay. \label{fig:fig4}}
\end{figure}

\section{Pseudopure state}\label{sec:sec4}
\subsection{Preparation method}\label{subsec:sec4a}
\indent  A labelled pseudopure state \cite{CPH98a,KCL98a} can be
prepared in this system using a combination of tailored RF pulses
and RF phase cycling (temporal averaging). A schematic of the
pulse sequence is shown in Figure~\ref{fig:fig4}. The first step
is a selective transfer of $^{1}$H polarization to the methylene
carbon $C_m$, utilizing the $^{1}$H/$^{13}$C coupling Hamiltonian
$\mathcal{H}_{CH}$.  While it is not necessary for this protocol
to begin with a \emph{selective} transfer, we demonstrate it here
because it may be used in other algorithms for controlled,
selective coupling of the qubit system to the bulk $^{1}$H system
(which can be considered as a thermal-bath of spin polarization).
To accomplish the transfer, a pulse sequence is applied on both
$^{1}$H and $^{13}$C channels synchronously, which, by design,
effectively removes $\mathcal{H}_{C}+\mathcal{H}_{H}$ from the
ensemble Hamiltonian \cite{CMG90a,Hae76a}. In such sequences,
homonuclear dipolar terms are refocussed by toggling the
interaction-frame Hamiltonian along the rotating-frame $\hat{x}$,
$\hat{y}$ and $\hat{z}$ directions, spending equal time along each
axis. Therefore, the dual sequence creates (to lowest order in the
Magnus expansion of the average Hamiltonian \cite{Hae76a}) an
effective $^{1}$H-$^{13}$C exchange Hamiltonian
\begin{equation}
\mathcal{H}_{\text{eff}}=\sum_{j\in C, k\in
H}\frac{d_{jk}}{3}\frac{(Z^{j}Z^{k}+Y^{j}Y^{k}+X^{j}X^{k})}{2},
\end{equation}
where $d_{jk}$ are pairwise $^{1}$H-$^{13}$C dipolar coupling
constants, and the indices $j,k$ run over all $^{13}$C,$^{1}$H
spins, respectively. In the special case that there is only one
coupled carbon/proton pair with a coupling of $d_{CH}$,
application of the sequence for a time $\tau=3/(4d_{CH})$ will
result in a state exchange between the two nuclei. Since this is
approximately the case for the strong methylene $^{1}$H-$^{13}$C
coupling in our oriented malonic acid system (see Table I), we may
implement a nearly selective polarization transfer to $C_m$ of
magnitude $P\simeq P_H=3.98 P_C$ ($P_C$ and $P_H$ are the thermal
equilibrium carbon and proton polarizations, respectively).
Furthermore, this selective transfer removes a very small amount
of polarization from the $^{1}$H bath, since only the methylene
protons on a dilute fraction of $^{13}$C-labelled molecules lose
their polarization. The remaining bulk $^{1}$H polarization is
preserved since the sequence is effectively a time-suspension
sequence for the bulk spins. In our system, a selective
polarization transfer of about $83\%$ of $P_H$ to $C_m$ was
achieved with a 12-pulse subsequence \footnote{The 12-pulse
subsequence does not fully average away the Zeeman interaction
even to $0^{th}$-order, so it is not strictly an
evolution-suspension sequence.  On the other hand, the duration of
the transfer is short enough that the evolution operator of the
bulk spin system is very close to the identity operator.} of the
Cory 48-pulse sequence \cite{CMG90a}; the duration of the 12-pulse
sequence was $40 \mu$s for maximum transfer.  The thermal
equilibrium $^{13}$C polarization can be removed prior to such a
transfer by rotating the equilibrium $^{13}$C state ($\propto Z$)
into the transverse ($\hat{x}-\hat{y}$) plane and allowing it to
dephase under local
$^{1}$H dipolar fields.\\
\indent Product operator terms denoting qubit states are ordered
as $C_1\otimes C_2\otimes C_m$; for example, the symbol $\id X Z$
corresponds to the state $\id^{1}\otimes X^{2}\otimes Z^{m}$.
Also, single-spin terms such as
$X^{1}\otimes\id^{2}\otimes\id^{m}$ are sometimes abbreviated as
$X^{1}$, for example. The polarization transfer is followed by a
$^{13}$C modulating RF pulse that transforms the state $\id\id Z$
to $ZZZ$, and then a collective $\pi/2$ pulse that rotates this to
$XXX$. In addition to single-quantum (1Q) terms, this state
contains the triple-quantum (3Q) coherence
$\sigma_{+}\sigma_{+}\sigma_{+}+\sigma_{-}\sigma_{-}\sigma_{-}$,
where $\sigma_{+}=X+iY$ and $\sigma_{-}=X-iY$.  A subsequent
modulating pulse (denoted 'TQPP' for 'triple-quantum to
pseudopure') transforms the 3Q coherence into the labelled
pseudopure state $X(\id+Z)(\id+Z)$. This state is observable as a
single NMR transition of $C_1$ \footnote{This is only
approximately true; a small amount of the pseudopure signal will
be found on other spin transitions due to the strong coupling
effect.  In this experiment, such weak signals are not separable
from the noise, although the effect is taken into account when
spectrally fitting the data.}. In order to cancel all other
signals arising from the 1Q terms in the state $XXX$, phase
cycling is used which exploits the $n$-proportional phase
acquisition of an $n$-quantum state under $\hat{z}$-axis rotation.
Choosing desired coherence pathways using phase cycling is widely
practiced in modern NMR spectroscopy \cite{EB87a}. By shifting the
phase of the RF by $\phi$ during the TQPP pulse, the unitary
transformation generated by the pulse is rotated by $\phi$ about
$Z^{\prime}=Z^{1}+Z^{2}+Z^{m}$:
\begin{equation}
U_{tqpp}(\phi)=R_{z}^{\phi}U_{tqpp}R_{z}^{-\phi},
\end{equation}
where $R_{z}^{\phi}\equiv e^{-i(Z^{\prime})\phi/2}$. Note this is
only true since the internal Hamiltonian of the system commutes
with $Z^{\prime}$, hence a $\hat{z}$-axis rotation can be
accomplished by acting only on the phase of the RF pulse.  We may
decompose the state prior to the TQPP pulse into 3Q and 1Q terms,
e.g. $XXX=\rho_{\pm 3Q}+\rho_{\pm 1Q}$. The final density matrix
is calculated as
\begin{align}
&\rho_{f}(\phi)=R_{z}^{\phi}U_{tqpp}R_{z}^{-\phi}(\rho_{\pm
3Q}+\rho_{\pm 1Q})R_{z}^{\phi}U^{\dag}_{tqpp}R_{z}^{-\phi}\nonumber\\
&=R_{z}^{\phi}((e^{-3i\phi}\sigma_{+}+e^{3i\phi}\sigma_{-})(\id+Z)(\id+Z)/8+e^{\mp
i\phi}\rho^{\prime})R_{z}^{-\phi},
\end{align}
where $\rho^{\prime}=U_{tqpp}(\rho_{\pm1Q})U^{\dag}_{tqpp}$.
Incrementing $\phi$ in units of $\frac{\pi}{3}$ for $6$ scans,
alternately adding and subtracting, adds constructively the 3Q
terms while cancelling the 1Q terms, since
\begin{equation}
\sum_{k=1}^{6}(-1)^{k-1}e^{\pm
nik\frac{\pi}{3}}=6\cdot\delta(n-3m),
\end{equation}
where $m$ is an odd integer, and here $m=1$ since $n\leq 3$. The
remaining $\hat{z}$-rotation $R_{z}^{\phi}$ is undone by
incrementing the phase of the receiver along with that of the TQPP
modulating RF pulse. The labelled pseudopure state is thus
obtained as
\begin{equation}
\frac{1}{6}\sum_{k=1}^{6}(-1)^{k-1}R_{z}^{-k\pi/3}\rho_{f}(k\pi/3)R_{z}^{k\pi/3}=X(\id+Z)(\id+Z)/4.
\end{equation}
The amount of signal contained in the resulting pseudopure state
is $1/4$ that of the input state, equal to the proportion of the
3Q part of the state $XXX$.

\begin{figure}
\includegraphics[height=7cm]{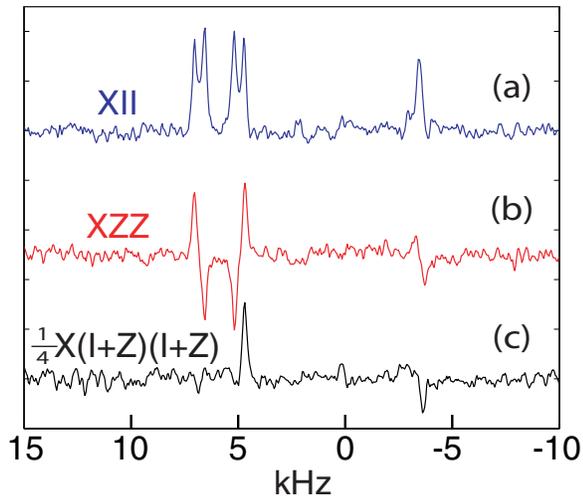}
\caption{Experimental results for the pseudopure state protocol.
(a) readout of the state $\id\id Z$ just after the selective
polarization transfer, with readout consisting of a swap from
$C_m$ to $C_1$ and collective $\pi/2$ pulse about the
$\hat{y}$-axis to produce $X\id\id$; (b) readout of the state
$ZZZ$ by a $C_1$-selective $\pi/2$ rotation to produce the
observable state $XZZ$; (c) the labelled pseudopure state yields a
single absorption peak from the $C_1$ multiplet. Fitting the data
by spectral simulation indicates a protocol state-correlation of
$87\%\pm 5\%$, using spectrum (a) as a reference.
\label{fig:fig5}}
\end{figure}

\subsection{Pseudopure state results}\label{subsec:sec4b}
Results for the pseudopure state protocol of Fig.~\ref{fig:fig4}
are shown in Fig.~\ref{fig:fig5}. It shows (a) readout of the
state $\id\id Z$ just after the selective polarization transfer,
with readout consisting of a state-swap from $C_m$ to $C_1$
followed by a collective $\pi/2$ pulse to produce $X\id\id$; (b)
readout of the state $ZZZ$ by a $C_1$-selective $\pi/2$ rotation
to produce the observable state $XZZ$; (c) the labelled pseudopure
state yields a single absorption peak from the $C_1$ multiplet.
Using the spectrum (a) as reference, the state-correlation of the
pseudopure state protocol, determined by spectral fitting, is
$87\%\pm 5\%$. Similarly, the state-correlation of the TQPP pulse,
using the $XZZ$ spectrum (b) as a reference, is $97\%\pm 5\%$.
These results serve as a benchmark for quantum control in the
strongly-coupled dilute molecular crystal system. They suggest
that dephasing due to Zeeman shift dispersion is largely
suppressed by these pulses, since the average free-induction
dephasing time of the qubits is $T^{\ast}_2\simeq 2$ ms, and the
total duration of the two modulating pulses, $\simeq 1.6$ ms, is a large fraction of that time.\\

\begin{figure}
\includegraphics[height=7.5cm]{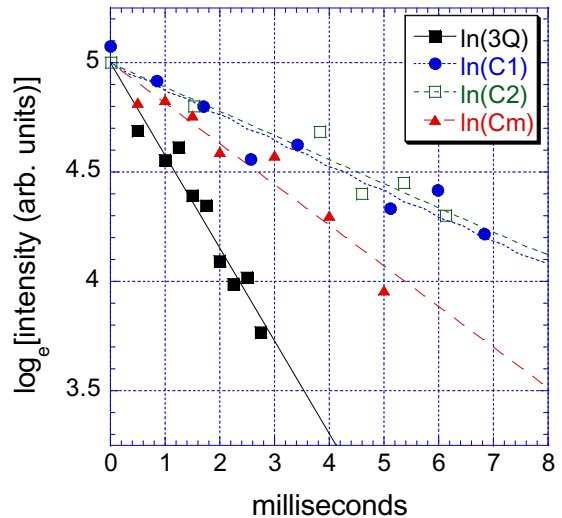}
\caption{Triple-quantum (3Q) and single-quantum dephasing data,
plotted as the natural log of the signal intensity. Exponential
fits indicate decay time constants (in milliseconds) of [$2.37\,
(3Q),\, 5.37 \,(C_m), \,9.07\, (C_2),\, 8.66 \,(C_1)$].
\label{fig:fig6}}
\end{figure}

\subsection{Dipolar dephasing of triple-quantum coherence}\label{subsec:sec4c}
\indent Dephasing of the 3Q coherence state
$\sigma_{+}\sigma_{+}\sigma_{+}+\sigma_{-}\sigma_{-}\sigma_{-}$
was measured by inserting a variable time delay $\tau(3Q)$
following the 'Y90' pulse in Fig.~\ref{fig:fig4}. A collective
$\pi$ pulse was also inserted in the center of $\tau(3Q)$ to
refocus Zeeman Hamiltonian terms (Hahn echo), so that the measured
dephasing rate reflects only the perturbing dipolar fields
experienced by the nuclei. This 3Q state is significant from a QIP
perspective since it consists of the two most fragile density
matrix elements in the three-qubit system, the extreme
off-diagonal elements that are contained in the so-called
"cat-state" $|000\rangle+|111\rangle$. The resulting signal decay
is shown in Fig.~\ref{fig:fig6} along with the 1Q dephasing data
for each qubit, also measured via Hahn echo. The 1Q decay data
were measured by first preparing the states $X^{1}$, $X^{2}$ and
$X^{m}$, and using the pair of outer spectral peaks from each spin
multiplet to gauge the signal, observing only at delays
corresponding to multiples of the dipolar oscillation periods of
these peaks. Exponential fits yield decay time constants (in
milliseconds) [$2.37\, (3Q),\, 5.37 \,(C_m), \,9.07\, (C_2),\,
8.66 \,(C_1)$], with experimental uncertainty $\simeq \pm 11\%$
for each value.  Within error, the observed 3Q dephasing rate is
equal to the sum of the three 1Q rates.  In the next section, we
will see that intermolecular $^{13}$C-$^{13}$C dipolar fields lead
to an approximately Lorentzian frequency broadening of the 1Q
transitions.  In this regime, the 1Q dephasing appears as a
Markovian process, and therefore we do not expect the 3Q rate to
carry any information about correlations in the 3Q dephasing. The
fact that the observed 3Q rate is the sum of the 1Q rates is
consistent with this picture.\\
\indent Finally, we remark that the faster 1Q dephasing of the
methylene carbon $C_m$ evident in Fig.~\ref{fig:fig6} is probably
due to a residual interaction with its neighboring proton, as the
methylene proton pair are strongly coupled ($22$ kHz) which makes
it difficult to remove the $19$ kHz $C_m$-$H_{m1}$ coupling under
the standard TPPM decoupling used here. This effect is directly
evident in the data presented in section~\ref{subsec:sec5b} (note,
however, the data in section~\ref{subsec:sec5b} was obtained in a
different sample at a slightly different orientation
with respect to the external field).\\

\section{Characterization II: Intermolecular dipolar fields}\label{sec:sec5}
\subsection{Modelling intermolecular dipolar dephasing}\label{subsec:sec5a}
\begin{figure}
\includegraphics[height=6cm]{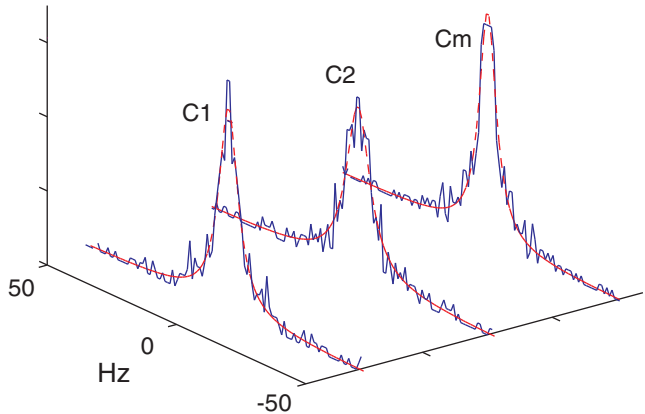}
\caption{Simulation of line-shapes resulting from intermolecular
dipolar couplings in $1.9\%$ molecular dilution $^{13}$C malonic
acid.  The simulated data (solid lines) are shown with Lorentzian
fits (dashed lines). The resulting linewidths (FWHM) for $C_1$,
$C_2$ and $C_m$ are $9$, $11$ and $8$ Hz respectively. Correcting
for the larger spin concentration due to presence of natural
abundance spins, the linewidths should be approximately $60\%$
broader. \label{fig:fig7}}
\end{figure}
\indent  We now turn our attention to characterizing
$^{13}$C-$^{13}$C intermolecular dipolar interactions in the
dilute malonic acid system. The general Hamiltonian of the system
is
\begin{equation}
\mathcal{H}=\sum_{j<k}d_{jk}h_{jk}+\sum_{l}\frac{\nu_{l}}{2}Z^{l},\label{eq:inter1}
\end{equation}
where $d_{jk}=d_{jk}(r_{jk},\theta_{jk})$ are the intermolecular
dipolar coupling constants that depend on the internuclear vector
of length $r_{jk}$ and orientation $\theta_{jk}$ (with respect to
the external field), $h_{jk}$ is the spin-operator of the form of
the dipolar coupling as in Eq.~\ref{eq:dip}, and $\nu_{l}$ are the
chemical shifts. Taking a reference spin $l=1$ to be a $C_1$ spin,
we may transform to the $C_1$ rotating frame by setting all
$\nu_{l}=0$ for $l\in\{C_1\}$, and setting $\nu_{l'}$ to the
correct offset frequencies for $l'\in\{C_2,C_3\}$, respectively.
The latter offset frequencies $\nu_{l'}$ are on the order of a few
kHz (see Table~\ref{tab:table1}), whereas the intermolecular
couplings $d_{jk}$ are much weaker ($<\!100$ Hz) due to the
$r_{jk}^{-3}$ scaling of the dipolar interaction.  We may
therefore truncate couplings between unlike spins
$C_{\alpha}$-$C_{\beta}$ to the heteronuclear form, and rewrite
the dipolar terms as an ensemble of reference-spin Hamiltonians
(e.g. for the $C_1$ reference ensemble):
\begin{align}
\mathcal{H}_{dip}&=\sum_{j,k}\mathcal{H}^{jk}_{dip};\nonumber\\
\mathcal{H}^{jk}_{dip}&=\frac{d_{\alpha jk}}{2}Z^{1j}Z^{\alpha jk}+\frac{d_{\beta jk}}{2}Z^{1j}Z^{\beta jk}\nonumber\\
&+\frac{d_{\gamma jk}}{4}(2Z^{1j}Z^{\gamma jk}-X^{1j}X^{\gamma
jk}-Y^{1j}Y^{\gamma jk})\label{eq:dip2},
\end{align}
where $j$ indicates a particular reference spin from the $C_1$
ensemble, and $k$ denotes the set of spin-labelled molecules that
interact with the $j^{th}$ reference spin.  $\alpha$, $\beta$ and
$\gamma$ indicate spins belonging to the $C_m$, $C_2$ and $C_1$
ensembles, respectively.  Note that the above Hamiltonian does not
include all dipolar terms in Eq.~\ref{eq:inter1}, since it only
includes strong-coupling terms between $C_1$ spins, and so
implicitly neglects much of the spin-diffusion dynamics. We wish
to make an estimate of the dephasing rate (i.e. line broadening)
of the reference spins due to this ensemble interaction
Hamiltonian (see \cite{NL57a} for seminal work along these lines).
An approximate result can be obtained in the spin-dilute regime by
assuming that the reference spin only interacts with one nearby
spin-labelled molecule, so that there is only one $k$ value
and $\mathcal{H}^{jk}_{dip}=\mathcal{H}^{j}_{dip}$. \\
\indent Under this restricted model, we will now describe the
dephasing process in terms of the quantum evolution of the
ensemble. The system consists of the reference spin, denoted
$\xi$, and the three interacting spins $\alpha,\beta,\gamma$. The
initial state is described by the density matrix
$\rho_{\xi}\otimes \rho_{\alpha\beta\gamma}$, where at thermal
equilibrium $\rho_{\alpha\beta\gamma}=\id$ is appropriate for
high-temperature NMR. The unitary operator acting on the $j^{th}$
ensemble member of the system at time $t$ is
\begin{equation}
U_{j}(t)=e^{-i \mathcal{H}^{j}_{dip} 2\pi t},
\end{equation}
The operators that act on the reduced density matrix $\rho_{\xi}$
of the reference spin, i.e. the Kraus operators, are derived from
the submatrices of $U$:
\begin{equation}
U_{mn}=\langle\psi^{m}_{\alpha\beta\gamma}|U|\psi^{n}_{\alpha\beta\gamma}\rangle,
\end{equation}
where $|\psi^{n}_{\alpha\beta\gamma}\rangle$ is the $n^{th}$
eigenvector of the $\alpha,\beta,\gamma$ system in some
eigenbasis. Since $\rho_{\alpha\beta\gamma}=\id$, the Kraus
operators $A_{m}$ are obtained by summing with equal weights over
the basis vectors:
\begin{equation}
A_{m}=2^{-q}\sum^{2^{q}}_{n=1}U_{mn},
\end{equation}
where the number of interacting spins is $q=3$ in our case. In the
standard computational basis,
$|\psi^{n}_{\alpha\beta\gamma}\rangle\in\{|000\rangle,|001\rangle,...,|111\rangle\}$,
we obtain for $U_{j}(t)$ operators of the form (subscripts
$k,l,m\in\{0,1\}$ denote the basis vector of the
$\alpha,\beta,\gamma$ system):
\begin{widetext}
\begin{align}
&A^{j}_{klm}(t)=\frac{1}{8}\times\nonumber\\
&[|0\rangle\langle0|e^{i\frac{\theta_{\alpha j}}{2}(-1)^{k}+i\frac{\theta_{\beta j}}{2}(-1)^{l}+i\frac{\theta_{\gamma j}}{2}(-1)^{m}}(\frac{e^{i\frac{\theta_{\gamma j}}{2}\delta_{m1}}+e^{-i\frac{\theta_{\gamma j}}{2}\delta_{m1}}}{2})+|1\rangle\langle1|e^{-i\frac{\theta_{\alpha j}}{2}(-1)^{k}-i\frac{\theta_{\beta j}}{2}(-1)^{l}-i\frac{\theta_{\gamma j}}{2}(-1)^{m}}(\frac{e^{i\frac{\theta_{\gamma j}}{2}\delta_{m0}}+e^{-i\frac{\theta_{\gamma j}}{2}\delta_{m0}}}{2})\nonumber\\
&+|1\rangle\langle0|\delta_{m0}e^{i\frac{\theta_{\alpha
j}}{2}(-1)^{k}+i\frac{\theta_{\beta
j}}{2}(-1)^{l}-i\frac{\theta_{\gamma
j}}{2}}(\frac{-e^{i\frac{\theta_{\gamma
j}}{2}}+e^{-i\frac{\theta_{\gamma
j}}{2}}}{2})+|0\rangle\langle1|\delta_{m1}e^{i\frac{\theta_{\alpha
j}}{2}(-1)^{k}+i\frac{\theta_{\beta
j}}{2}(-1)^{l}-i\frac{\theta_{\gamma
j}}{2}}(\frac{-e^{i\frac{\theta_{\gamma
j}}{2}}+e^{-i\frac{\theta_{\gamma j}}{2}}}{2})]
\end{align}
\end{widetext}
where $\theta_{\eta j}=2\pi t d_{\eta j}$, and $\delta_{ab}$ is the Kronecker delta.\\
\indent To study dephasing behavior, we apply the eight Kraus
operators $A^{j}_{klm}(t)$ to the reference spin state
$\rho_{\xi}=X=|0\rangle\langle1|+|1\rangle\langle0|$, obtaining
\begin{equation}
\rho^{j}_{\xi}=\sum_{klm}A^{j}_{klm}\rho_{\xi}(A^{j}_{klm})^{\dag}=
\sum_{klm}(a^{j}_{klm}|0\rangle\langle1|+b^{j}_{klm}|1\rangle\langle0|),
\end{equation}
where
\begin{align}
&a^{j}_{klm}=\frac{1}{8}\times\nonumber\\
&e^{i\theta_{\alpha j}(-1)^{k}+i\theta_{\beta
j}(-1)^{l}}\frac{e^{i\theta_{\gamma
j}((-1)^{m}+1/2)}+e^{i\theta_{\gamma
j}((-1)^{m}-1/2)}}{2}\label{eq:a}
\end{align}
and
\begin{align}
&b^{j}_{klm}=\frac{1}{8}\times\nonumber\\
&e^{-i\theta_{\alpha j}(-1)^{k}-i\theta_{\beta
j}(-1)^{l}}\frac{e^{-i\theta_{\gamma
j}((-1)^{m}-1/2)}+e^{-i\theta_{\gamma
j}((-1)^{m}+1/2)}}{2}\label{eq:b}
\end{align}
Inspection of equations~\ref{eq:a} and~\ref{eq:b} shows that
$\rho^{j}_{\xi}$ gains phases $\pm \theta_{\alpha j}\pm
\theta_{\beta j}\pm \theta_{\gamma j}/2$ and $\pm \theta_{\alpha
j}\pm \theta_{\beta j}\pm 3\theta_{\gamma j}/2$ characteristic of
generic binomial distributions. Averaging over the ensemble of
Hamiltonians $\mathcal{H}^{j}_{dip}$, we obtain
$\rho_{\xi}(t)=\sum^{N}_{j=1}\rho^{j}_{\xi}(t)/N$. Defining a
correlation function $F_{x}(t)=Tr(\rho_{\xi}(t)\cdot X)$, its
frequency spectrum is given by the Fourier transformation
$\hat{F}_{x}(\omega)=\int^{\infty}_{-\infty}e^{i \omega
t}F_{x}(t)$. Equations~\ref{eq:a} and~\ref{eq:b} make clear that
$\hat{F}_{x}(\omega)$ will simply reflect the ensemble
distribution of the intermolecular coupling constants leading to
frequencies $2\pi\{\pm d_{\alpha j}\pm d_{\beta j}\pm d_{\gamma
j}/2\}$ and $2\pi\{\pm d_{\alpha j}\pm
d_{\beta j}\pm 3 d_{\gamma j}/2\}$.\\
\indent To make a concrete calculation of the distribution of
coupling constants $d_{\eta j}$, let us further assume that the
interacting molecule lies within the first shell of 26 neighboring
unit cells. Each unit cell contains two (magnetically equivalent)
molecules, giving a total of $156$ atomic sites ($159$ after
adding the 3 atomic sites of the unit cell neighbor to the
reference molecule). We calculated the coupling constants to each
of these sites using the Cartesian atomic coordinates obtained by
x-ray diffraction \cite{JRS94a} and the unit cell vectors
$(\vec{A},\vec{B},\vec{C})$ (see Fig.~\ref{fig:fig2}).  The
couplings are explicitly of the form:
\begin{equation}
d_{k}(r_{k},\theta_{k})=\gamma^{2}_C\hbar\frac{1-3cos^2(\theta_{k})}{2r_{k}^3},
\end{equation}
where $\gamma_C$ is the gyromagnetic ratio of $^{13}$C, $r_{k}$ is
the length of the internuclear vector connecting spin $k$ to the
reference spin, and $\theta_{k}$ is the angle between this vector
and the external magnetic field direction.  Note that since there
are $53$ molecular sites we are considering, the random occupation
of one site corresponds to a labelled-molecule concentration of
$\frac{1}{53}=1.9\%$, which is in the range of our sample
concentrations. Following the discussion above, we constructed
frequency histograms by averaging the spectral frequencies
$2\pi\{\pm d_{\alpha j}\pm d_{\beta j}\pm 3 d_{\gamma j}/2\}$ over
the distribution of coupling constants to the 53 molecular sites.
The same procedure was carried out for $C_1$, $C_2$ and $C_m$ as
reference spins. Figure~\ref{fig:fig7} shows the resulting
histograms calculated for the crystal orientation used throughout.
Lorentzian functions provide good fits for the purpose of
determining the spectral linewidths, as expected for a
dilute spin system \cite{Abr61a}.\\
\indent The simulated spectra of figure~\ref{fig:fig7} correspond
to the expected indirect-dimension line-shapes that would result
in a two-dimensional NMR experiment under a Hahn echo sequence
\cite{EB87a}.  The fits yield full-width half-maximum (FWHM)
linewidths $\Delta\nu\sim 10$ Hz. The presence of single $^{13}$C
spins due to natural abundance ($1.1\%$ atomic concentration) adds
to the total spin concentration. This broadens the linewidth
estimate by a factor $\simeq \frac{\eta+1.1\%}{1.9\%}$ for a
dilute labelled-molecule percentage $\eta<10\%$, since linewidth
is approximately linear in spin concentration in this dilute
regime \cite{Abr61a}.  To account for interactions with molecules
beyond the first neighboring unit cells in this static broadening
picture (i.e. ignoring spin-diffusion dynamics), we can make a
crude shell-model approximation. The broadening from the $n$ spins
in a spherical shell of radius $R$, $n(R)\propto R^{2}$, will go
as $R^{-3}\sqrt{n(R)}\propto R^{-2}$.  This is true since the
width (i.e. standard deviation) of a binomial distribution scales
as $\sqrt{n}$ and the dipolar interaction scales as $R^{-3}$.
Hence, the linewidths should be larger by a factor
$\sum^{\infty}_{k=1}k^{-2}=\pi^{2}/6$. The results, adjusted for
$\eta=1.6\%$, are summarized in Table~\ref{tab:table2} along with
experimental data obtained in a $1.6\%$ labelled-molecule dilution
crystal. The experimental values indicate effective Lorentzian
linewidths of the natural abundance $^{13}$C spins measured via
Hahn echo, and by a multiple-pulse sequence consisting of the
MREV-8 dipolar refocusing sequence \cite{Hae76a} in combination
with the Hahn echo (detailed in the following section). In the
table, 'Simulation II' refers to the aforementioned spectral
simulations. These values should correspond to the Hahn echo
experimental data. 'Simulation I' refers to simulations in which
like-spin $C_{\alpha}$-$C_{\alpha}$ couplings were omitted. These
values should correspond roughly with the dipolar+Hahn refocusing data, as described in the next section.\\

\begin{table*}
\caption{\label{tab:table2} Simulated and experimental dephasing
rates for natural abundance spins in $1.6\%$ isotopic dilution
crystal, in terms of Lorentzian linewidths (all in Hz).
Simulations I/II refer to the exclusion/inclusion of couplings
between indistinguishable spins. 'Simulation II' values should be
compared to the Hahn echo results, and 'Simulation I' values
should be compared to the dipolar+Hahn refocusing results. Under
the latter sequence, effective linewidths are reduced by nearly
two orders of magnitude compared to the free-induction decay
spectrum, with the exception of $C_{m}$, whose coherence time is
limited by residual proton dephasing (see footnote $a$).}
\begin{tabular}{|>{\centering}m{0.5in}|>{\centering\footnotesize}m{0.75in}|>{\centering\footnotesize}m{0.75in}|>{\centering\footnotesize}m{0.75in}|>{\centering\footnotesize}m{0.75in}|>{\footnotesize}m{0.5in}|}
 \hline
 \multirow{2}{1in}{\centering\footnotesize{}}&\multicolumn{5}{c|}{Dipolar and static-field dephasing (1.6$\%$ crystal)} \\\cline{2-6}
 & Simulation I &  dipolar + Hahn refocusing & Simulation II & Hahn echo & FID\\ \hline
 \centering\footnotesize
 C1 & 6.7 & $\sim$ 2 & 20.1 & 25 & 133 \\ \hline
  \centering\footnotesize
 C2 & 7.6 & $\sim$ 3 & 25.1 & 32 &  122 \\ \hline
  \centering\footnotesize
 Cm & 13.9 & 11/20\footnotemark[1] & 18.0 & 25 &  103 \\ \hline

 \hline
 \end{tabular}
 \footnotetext[1]{$11$ Hz and $20$ Hz correspond to proton decoupling powers of $330$ kHz and $270$ kHz,
 respectively.}
 \end{table*}

 \begin{figure}
\includegraphics[height=7.5cm]{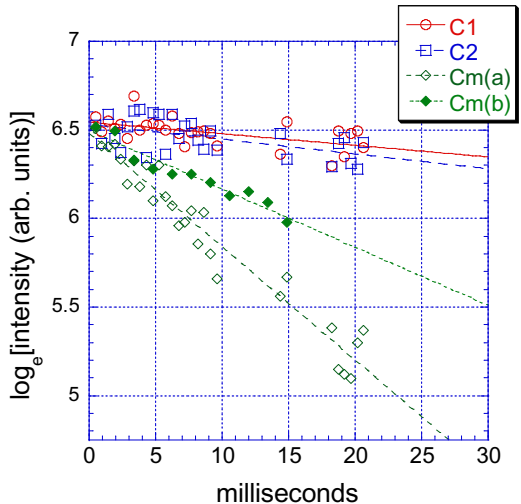}
\caption{Decay of signal intensity under the dipolar+Hahn
refocusing sequence for natural abundance spins in the $1.6\%$
isotopic dilution crystal. $C_{m}(a)$ and $C_{m}(b)$ had TPPM
proton decoupling RF power levels of $270$ kHz and $330$ kHz,
respectively. Exponential fits to the data indicate effective
coherence times (in milliseconds) of [$150$ ($C_{1}$), $120$
($C_{2}$), $16$ ($C_{m}(a)$), $30$ ($C_{m}(b)$)], although clearly
the values obtained for the slower $C_{1}$ and $C_{2}$ decays are
more approximate than those for $C_{m}$.  Dephasing times $>21$ ms
were not explored here due to the heating effects of high-power
proton decoupling on the room temperature probe/sample.
\label{fig:fig8}}
\end{figure}

\subsection{Coherence times under dipolar+Hahn refocusing}\label{subsec:sec5b}
\indent Application of a Zeeman and homonuclear dipolar refocusing
sequence was investigated, to compare with the simulations of the
previous section, and to explore the attainability of longer
$^{13}$C coherence times. The employed sequence is the MREV-8
\cite{Hae76a} dipolar decoupling sequence in combination with a
Hahn echo. The pulse spacing of the MREV-8 sequence was set such
that one eight-pulse cycle was $240\mu$s in duration. A single
$\pi$-pulse was applied in the center of a given dephasing period
to refocus the effective Zeeman field \cite{Hae76a} of the MREV-8
sequence (the effective field is in the $\hat{z}-\hat{x}$ plane,
so that a $\pi$-pulse about the $\hat{y}$-axis will refocus it).
It is a quasi-evolution-suspension sequence because it refocuses
like-spin dipolar evolution on a relatively short time-scale
compared to the refocusing of Zeeman evolution. Therefore it does
not fully remove effective dipolar couplings resulting from
unlike-spin $C_{\alpha}$-$C_{\beta}$ couplings (their magnitude
will be substantially scaled, however). The data are shown in
Figure~\ref{fig:fig8}. The decay of the magnetization of the
natural abundance spins is shown, since the aim here is to see the
effect of refocusing the \emph{inter}molecular dipolar couplings
on coherence times. Monitoring the natural abundance spins is
ideal in this regard, since the average intermolecular dipolar
environment is identical for all $^{13}$C spins in the sample, and
for the natural abundance spins, it is the \emph{only} $^{13}$C
dipolar environment. On the other hand, efficient refocusing of
the \emph{intra}molecular couplings would require a shorter MREV-8
sequence, and therefore more pulses for a given dephasing time,
yielding greater loss of signal due to pulse imperfections.  The
ability to suspend the evolution of the qubit spins (e.g. to
generate an effective propagator equal to the identity) is an
important benchmark for control \cite{LMY+05a} and will be explored in future work.  \\
\indent The coherence times of $C_1$ and $C_2$ indicated in
Fig.~\ref{fig:fig8} (and the corresponding Lorentzian linewidths
shown in Table~\ref{tab:table2}) demonstrate a $\sim 50$-fold
increase in coherence time compared with the free-induction
dephasing times (Table~\ref{tab:table1}). Further, the residual
broadening is comparable to the strength of the
$C_{\alpha}$-$C_{\beta}$ dipolar interactions, as expected. The
coherence time of the methylene carbon $C_m$ is clearly limited by
the efficiency of proton decoupling. We note that dephasing times
$>21$ ms were not explored due to the heating effects of
high-power proton decoupling on the room temperature probe/sample.
Comparing the Hahn echo and dipolar/Hahn echo data, it is seen
that Zeeman shift dispersion contributes $\sim 90$ Hz ($1.2$ ppm)
to the free-induction linewidths, whereas the intermolecular
dipolar fields contribute $\sim 25$ Hz in the $1.6\%$ dilution
crystal. Sequences that efficiently average all spin interactions,
such as the Cory 48-pulse sequence \cite{CMG90a}, could be used to
test the limits of line-narrowing.

\section{Discussion}\label{sec:sec6}
\indent In this paper, we have established benchmark results and
described a methodology for controlling solid-state NMR qubits
upon which future work can build and improve. High-fidelity
quantum control was demonstrated through the preparation of a
pseudopure state. That result, along with unitary simulations of
the fidelity for the 3-qubit controlled-(not$\otimes$not) gate,
suggests that ensemble inhomogeneities involving linear
Hamiltonian terms (Zeeman dispersion, RF amplitude) may be
suppressed by strongly modulating RF pulses. Additionally, the
method of strongly modulating pulses is well-suited to finding
solutions for generating arbitrary quantum gates, approaching
time-optimality, in systems with complex internal Hamiltonians.
Measurement of the triple-quantum dipolar dephasing rate was
carried out, as an application of the pseudopure state protocol.
Simulations of the intermolecular dipolar field spectrum predicted
dipolar linewidths in reasonable agreement with experimental
results obtained under appropriate refocusing sequences.
Refocusing of the Zeeman and like-spin dipolar terms lead to a
$50$-fold increase in ensemble coherence times, and suggests that
much narrower effective linewidths could be achieved using sequences that efficiently remove all spin interactions.\\
\indent These results demonstrate the potential for achieving
universal quantum control in solid-state NMR systems with the high
precision required for quantum computation and other information
processing tasks. Coupled with state purification techniques like
dynamic nuclear polarization, such control should enable a
powerful new QIP testbed reaching up to $\sim 20-30$ qubits
\cite{CLK+00a}. Furthermore, the solid-state setting is more
general than the liquid-state, for example, by allowing one to
couple the qubits to a bulk-spin heat bath \cite{CLK+00a}, as
demonstrated by the selective polarization transfer that was
employed to begin the pseudopure state protocol. Similar
techniques will enable implementation of quantum error correction
and heat-bath algorithmic cooling
\cite{SV90a} protocols, the latter having been already demonstrated in the malonic acid system \cite{BMR+05a}.\\
\indent To further improve control, the strongly modulating pulse
optimization procedure must be improved to yield simulated
fidelities comparable to those achievable in the liquid state
(that is, unitary fidelities in the range of $99-99.9\%$ as
opposed to $\sim 95-99\%$).  Moreover, the experimental system
must be carefully engineered so that the implemented control
fields are sufficiently faithful to the simulated pulses.
Extending to larger numbers of qubits ($\gtrsim\!12$) will likely
require some form of hybrid control that utilizes multiple-pulse,
average Hamiltonian techniques in combination with the numerical
optimization methods discussed. Additionally, scalable control
methods that have been successfully demonstrated in the
weak-coupling regime \cite{KLMT00a,NMR+05a} could potentially be
merged with the strongly modulating pulse methodology to achieve a scalable pulsefinder for
appropriately tailored systems that include strongly-coupled spins.\\
\indent In conclusion, we have characterized in detail a novel
solid-state NMR system for quantum information processing, and
used it to demonstrate high-fidelity quantum control.  This lays a
foundation for future studies in the malonic acid system, and in
larger systems, desirably with high polarization.

\begin{acknowledgments}
We wish to acknowledge NSERC, CIAR, ARDA, ARO and LPS for support;
M. Ditty, N. Taylor and W. Power for experimental assistance.
\end{acknowledgments}

\end{document}